\documentclass[twocolumn,showpacs,preprintnumbers,amsmath,amssymb]{revtex4}

\usepackage{CJK}
\usepackage{graphicx}
\usepackage{dcolumn}
\usepackage{bm}

\def\btbl{\begin{tabular}} \def\etbl{\end{tabular}}
\def\bcc{\begin{center}} \def\ecc{\end{center}}
\def\beq{\begin{equation}} \def\eeq{\end{equation}}
\def\btbl{\begin{tabular}} \def\etbl{\end{tabular}}

\def\E941{{\footnotesize E941}} \def\E864{{\footnotesize E864}}
\def\NA49{{\footnotesize NA49}} \def\NA35{{\footnotesize NA35}}

\begin{document}
\title{Effect of the Wood-Saxon nuclear distribution on the chiral magnetic field in Relativistic Heavy-ion Collisions}

\author{Yu-Jun Mo$^{1,2}$}
\author{Sheng-Qin ~Feng$^{1,2,3}$}
\author{Ya-Fei ~Shi$^1$}

\affiliation{$^1$College of Science, Three Gorges University,
Yichang 443002, China} \affiliation{$^2$Key Laboratory of
Quark and Lepton Physics (Huazhong Normal University), Ministry of
Education£¬Wuhan 430079£¬China} \affiliation{$^3$School of Physics
and Technology, Wuhan University, Wuhan 430072, China}

\begin{abstract}
The formation of the QCD vacuum with nonzero winding number $Q_w$ during relativistic
heavy-ion collisions breaks the parity and charge-parity symmetry.
A new kind of field configuration can separate charge in the presence of
a background magnetic field-the "chiral magnetic effect".
The strong magnetic field and the QCD vacuum can both completely be produced
in the noncentral nuclear-nuclear collision.
Basing on the theory of Kharzeev,Mclerran and Warringa, we use the Wood-Saxon nucleon distribution to replace
that of the uniform distribution to improve the magnetic field calculation method
of the noncentral collision. The chiral magnetic field distribution at LHC(Large Hadron Collider) energy regions are predicted.
We also consider the contributions to the magnetic field of the total charge given by the produced quarks. \\

\vskip0.2cm \noindent Keywowds: ~~Chiral magnetic effect,~Non-central collision, ~Wood-Saxon distribution
\end{abstract}

 \maketitle

\section{Introduction}
\label{intro}
When two heavy ions collide with a nonzero impact
parameter, a magnetic(electromagnetic) field of enormous
magnitude is created in the direction of angular momentum
of the collision~\cite{lab1,lab2,lab3}.
If a nonzero chirality is present in such a situation, an electromagnetic
current will be induced in the direction of the magnetic
field. This is the so-called chiral magnetic effect ~\cite{lab4,lab5,lab6}.

One of the most exciting signals of the deconfinement and the chiral phase transitions
in heavy-ion collisions, the chiral magnetic effect~\cite{lab7,lab8,lab9,lab10,lab11,lab12}, predicts the
preferential emission of charged particles along the direction of angular momentum
in the case of noncentral heavy-ion collisions due to the presence of nonzero
chirality. As it was stressed in Refs.~\cite{lab1,lab2,lab3}, both the deconfinement and the chiral phase transitions
are the essential requirements for the chiral magnetic effect to take place.

In a heavy-ion collision this current leads to an excess of
positive charge on one side of the reaction plane (the plane
in which the beam axis and the impact parameter lies) and
negative charge on the other; the resulting charge asymmetry
is also modulated by the radial flow and the transport
properties of the medium. This charge asymmetry can be
investigated experimentally~\cite{lab13,lab14,lab15,lab16,lab17,lab17} using the observable proposed~\cite{lab18,lab19,lab20,lab21}.

In recent years, Kharzeev, Mclerran and Warringa(KMW) presented new evidence of a charge-parity(CP) violation
in relativistic heavy-ion collisions caused by the nonzero $Q_{w}$ gauge field configurations
\cite{lab1,lab2}. KMW proposed that this kind of configuration can separate charge which means
the right- and left-hand quarks created during the collisions will move oppositely with
respect to the reaction plane in the presence of a background magnetic field.
Also, high energy physics experiments have obtained a series of results to support the chiral magnetic effect.

KMW~\cite{lab1} presented a novel mechanism for charge separation. The topological charge
changing transitions provide the parity(P)- and CP violations necessary for charge separation. The variance
of the net topological charge change is proportional to the total number of topological
charge changing transitions. Hence, if sufficiently hot matter is created in heavy-ion collisions
so that topological charge transitions can take place, we expect on average in each event a finite
amount of topological charge change.

Charge separation needs a symmetry axis along which the separation can take place. The
only symmetry axis in a heavy-ion collision is angular momentum which points in the direction
perpendicular to the reaction plane. In central collisions there is no symmetry axis, so in that case
charge separation should vanish. The strong magnetic field and the QCD vacuum can both completely be produced
in the non-central nuclear-nuclear collision.
Based on the theory of KMW, we use the Wood-Saxon nucleon distribution to replace
that of the uniform distribution to improve the magnetic field calculation method
of the noncentral collision. The chiral magnetic field distribution at LHC energy regions are predicted in this paper.

The paper is organized as follows. The modified calculation of chiral magnetic field and the comparison of our new results with
that of given by KMW  are
described in Sec. II, along with the predicted results of LHC energy region. The produced particle contribution to the magnetic field is considered  in Sec. III.
A summary is given in Sec. IV.

\section{The Modified calculation of Chiral Magnetic field}
The situation with the experimental search for the local
strong parity violation drastically changed once it was
noticed ~\cite{lab1,lab2,lab3,lab4,lab5,lab6} that in noncentral nuclear collisions it would
lead to the asymmetry in the emission of positively
and negatively charged particle perpendicular to the reaction
plane. Such a charge separation is a consequence of
the difference in the number of quarks with positive and
negative helicities positioned in the strong magnetic field of a noncentral nuclear collision, the so-called
chiral magnetic effect (CME).

We begin with a charged particle moving along the direction of the $z$ axis as shown in Fig. 1.
The magnetic field around it can be given by
\begin{equation}
\vec{B}=\frac{1}{c^2}\vec{v}\times\vec{E}
\label{eq:eq1} 
\end{equation}

If the movement is relativistic, at the time $t=0$, the charge is the origin of the coordinate. The magnitude of the magnetic field
$\vec{B}$ is given by
\begin{eqnarray}
B=\frac{1}{4\pi\varepsilon_0c^2}
\frac{qv(1-\beta^2)\sin\theta}{r^2(1-\beta^2\sin\theta)^{3/2}}.
\label{eq:eq2} 
\end{eqnarray}

Now we consider a particle with charge $Z$ and rapidity $Y$ traveling along the $z$ axis.
At $t=0$ the particle can be found at position $\vec{x}^{\prime}_{\bot}$; the magnetic field
at the position $\vec{x}=(\vec{x}_\bot,z)$ caused by the particle is given by

\begin{eqnarray}
\lefteqn{e\vec{B}(\vec{x})=Z\alpha_{EM}\sinh Y\times}\nonumber\\
&&\frac{(\vec{x}^{\prime}_{\bot}-\vec{x}_\bot)\times\vec{e}_z}
{[(\vec{x}^{\prime}_{\bot}-\vec{x}_\bot)^2+(t\sinh Y-z\cosh Y)^2]^{3/2}}
\label{eq:eq3} 
\end{eqnarray}

\begin{figure}[h!]
\centering \resizebox{0.52\textwidth}{!}{
\includegraphics{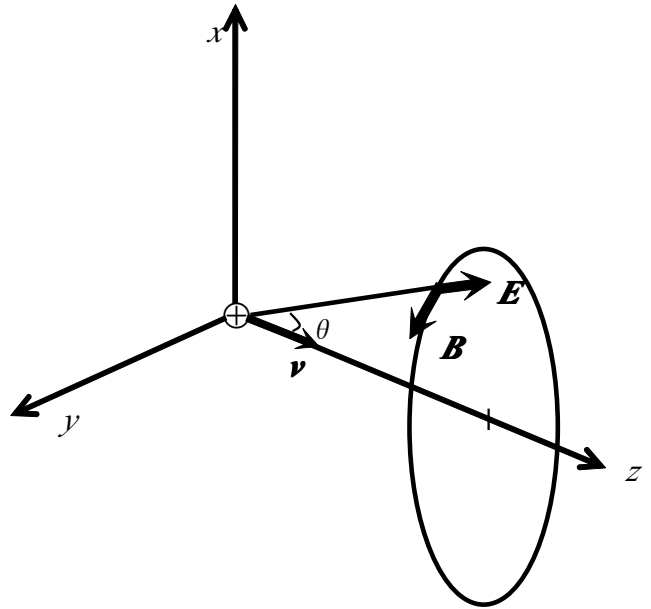}}
\caption{The magnetic field around a moving charged particle.}
\label{fig1}
\end{figure}

Now we suppose two similar nuclei with charge $Z$ and radius $R$ are traveling
in the positive and negative $z$ direction with rapidity $Y_0$.
At $t=0$ they have a noncentral collision with impact parameter $b$ at the origin point.
We take the center of the two nuclei at $x=\pm b/2$ at  time $t=0$
so that the direction of $b$ lies along the $x$ axis(see Fig.2).

As the nuclei are nearly traveling with the speed of light in typical heavy-ion collision experiments,
the Lorentz contraction factor $\gamma$ is so large that we can consider the two included nuclei as pancake shaped.
As a result, the nucleon's number density of each nuclei at $\vec{x}^{\prime}=(\vec{x}^{\prime}_{\bot},z)$
can be given by

\begin{eqnarray}
\rho_{s\pm}(\vec{x}^{\prime}_{\bot})=\frac{2}{4/3\pi R^3}
\sqrt{R^2-(\vec{x}^{\prime}_{\bot}\pm\vec{b}/2)^2}
\label{eq:eq4} 
\end{eqnarray}

\begin{figure}[h!]
\centering \resizebox{0.5\textwidth}{!}{
\includegraphics{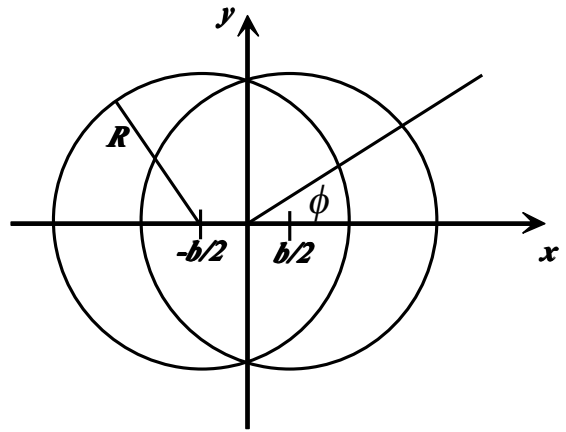}}
\caption{Cross-sectional view of a noncentral heavy-ion collision along the $z$ axis.
            The two nuclei have radii $R$, move in opposite directions, and collide with impact parameter $b$.
            The plane $y=0$ is called the reaction plane. The angle $\phi$ is an azimuthal angle with respect to the reaction plane.
            The region in which the two nuclei overlap contains the participants, the regions in which they do not overlap contain the spectators.}
\label{fig2}
\end{figure}

As a result, it seems that the nucleon distribution on average in a nucleus is an approximate result before considering
the Lorentz contraction. In Ref.[1], KMW model used the uniform nuclear distribution as the nuclear distribution. But for a real situation, the nucleon distribution is not strictly uniform.
It seems more reasonable to use the Wood-Saxon distribution replace the uniform
distribution. We use the Wood-Saxon distribution in this paper,

\begin{eqnarray}
n_A(r)=\frac{n_0}{1+\exp{(\frac{r-R}{d})}},
\label{eq:eq5} 
\end{eqnarray}

\noindent here $n_0$=0.17fm$^{-3}$, $d$=0.54fm and the radius $R$=1.12A$^{1/3}$fm. Considering the Lorentz contraction,
the density in the two-dimensional plane can be given by

\begin{eqnarray}
\rho_{\pm}(\vec{x}^\prime_\bot)=N\cdot\int_{-\rm R}^{\rm R}dz'\frac{n_0}{1+\exp(\frac{\sqrt{(x'\mp{b/2})^2+y'^{2}+z'^{2}}-{\rm R}}{d})},
\label{eq:eq6} 
\end{eqnarray}

\noindent where $N$ is the normalization constant. The number densities should be normalized as

\begin{eqnarray}
\int{d}\vec{x}^\prime_\bot\rho_{\pm}(\vec{x}^\prime_\bot)=1.
\label{eq:eq7} 
\end{eqnarray},

We now estimate the strength of the magnetic field at position $\vec{x}=(\vec{x}_\bot,z)$
caused by the two traveling nuclei. We are only interested in the time $t>0$, i.e. just after the collision.
Then we can split the contribution of particles to the magnetic field in the following way

\begin{eqnarray}
\vec{B}=\vec{B}^+_s+\vec{B}^-_s+\vec{B}^+_p+\vec{B}^-_p
\label{eq:eq8} 
\end{eqnarray}

\noindent where $\vec{B}^\pm_s$ and $\vec{B}^\pm_p$ are the the contributions of the spectators and the participants
moving in the positive or negative $z$ direction, respectively. For spectators, we assume that they do not scatter at all
and that they keep traveling with the beam rapidity $Y_0$. According to Eq.(3), we use the density above and find

\begin{eqnarray}
\lefteqn{e\vec{B}^\pm_s(\tau,\eta,\vec{x}_\bot)=\pm Z\alpha_{EM}\sinh(Y_0\mp\eta)
\int{d}^2\vec{x}^\prime_\bot\rho_{\pm}(\vec{x}^\prime_\bot)}\nonumber\\
&&\times[1-\theta_\mp(\vec{x}^\prime_\bot)]\frac{(\vec{x}^\prime_\bot-\vec{x}_\bot)\times\vec{e}_z}
{[(\vec{x}^\prime_\bot-\vec{x}_\bot)^2+\tau^2\sinh(Y_0\mp\eta)^2]^{3/2}},
\label{eq:eq9} 
\end{eqnarray}

\noindent where $\tau=(t^2-z^2)^{1/2}$ is the proper time , $\eta=\frac{1}{2}\ln[(t+z)/(t-z)]$ is the space-time rapidity, and

\begin{eqnarray}
\theta_\mp(\vec{x}^\prime_\bot)=\theta[R^2-(\vec{x}^\prime_\bot\pm\vec{b}/2)^2].
\label{eq:eq10} 
\end{eqnarray}

Here, We would like to  neglect the contribution of the production particles created by the interactions approximately
and so we just need to take into account the contribution of the participants that were originally there.
The distribution of participants that remain traveling along the beam axis is given by

\begin{eqnarray}
f(Y)=\frac{a}{2\sinh(aY_0)}{\rm e}^{aY},  \hskip1cm -Y_{0}\leq{Y}\leq{Y_{0}}
\label{eq:eq11} 
\end{eqnarray}

\noindent Experimental data shows that $a\approx1/2$, consistent with the baryon junction stopping mechanism. The contribution
of the participants to the magnetic field can be also given by

\begin{eqnarray}
e\vec{B}^\pm_p(\tau,\eta,\vec{x}_\bot)=\pm Z\alpha_{EM}\int{\rm d}^2\vec{x}^\prime_\bot
\int{\rm d}Y f(Y)\sinh(Y\mp\eta)\nonumber\\
\times\rho_{\pm}(\vec{x}^\prime_\bot)\theta_\mp(\vec{x}^\prime_\bot)
\frac{(\vec{x}^\prime_\bot-\vec{x}_\bot)\times\vec{e}_z}
{[(\vec{x}_\bot^\prime-\vec{x}_\bot)^2+\tau^2\sinh(Y\mp\eta)^2]^{\frac{3}{2}}}
\label{eq:eq12} 
\end{eqnarray}

We calculate the magnetite of the magnetic field at the origin ($\eta=0,\vec{x}_\bot=0$) in which case
it is pointing in the $y$ direction. We took a Au-Au collision with different beam rapidities
and different impact parameters.

\begin{figure}[h!]
\centering \resizebox{0.5\textwidth}{!}{
\includegraphics{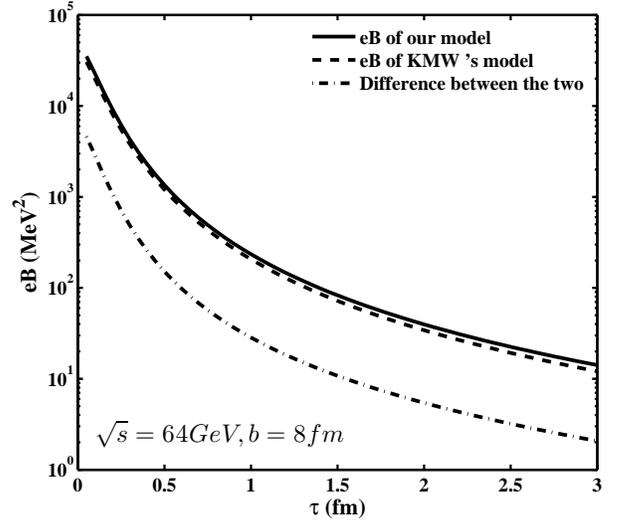}}
\caption{The dependence of the magnetic field on proper time for Au-Au collisions with $\sqrt{s}$ = 64 GeV and $b$ = 8 fm.
The solid line denotes our calculation results by using the Wood-Saxon nuclear distribution, the dashed line is the KMW results by using
the uniform nuclear distribution, and the dash dotted line denotes the difference  between our model and KMW's model.}
\label{fig3}
\end{figure}

\begin{figure}[h!]
\centering \resizebox{0.5\textwidth}{!}{
\includegraphics{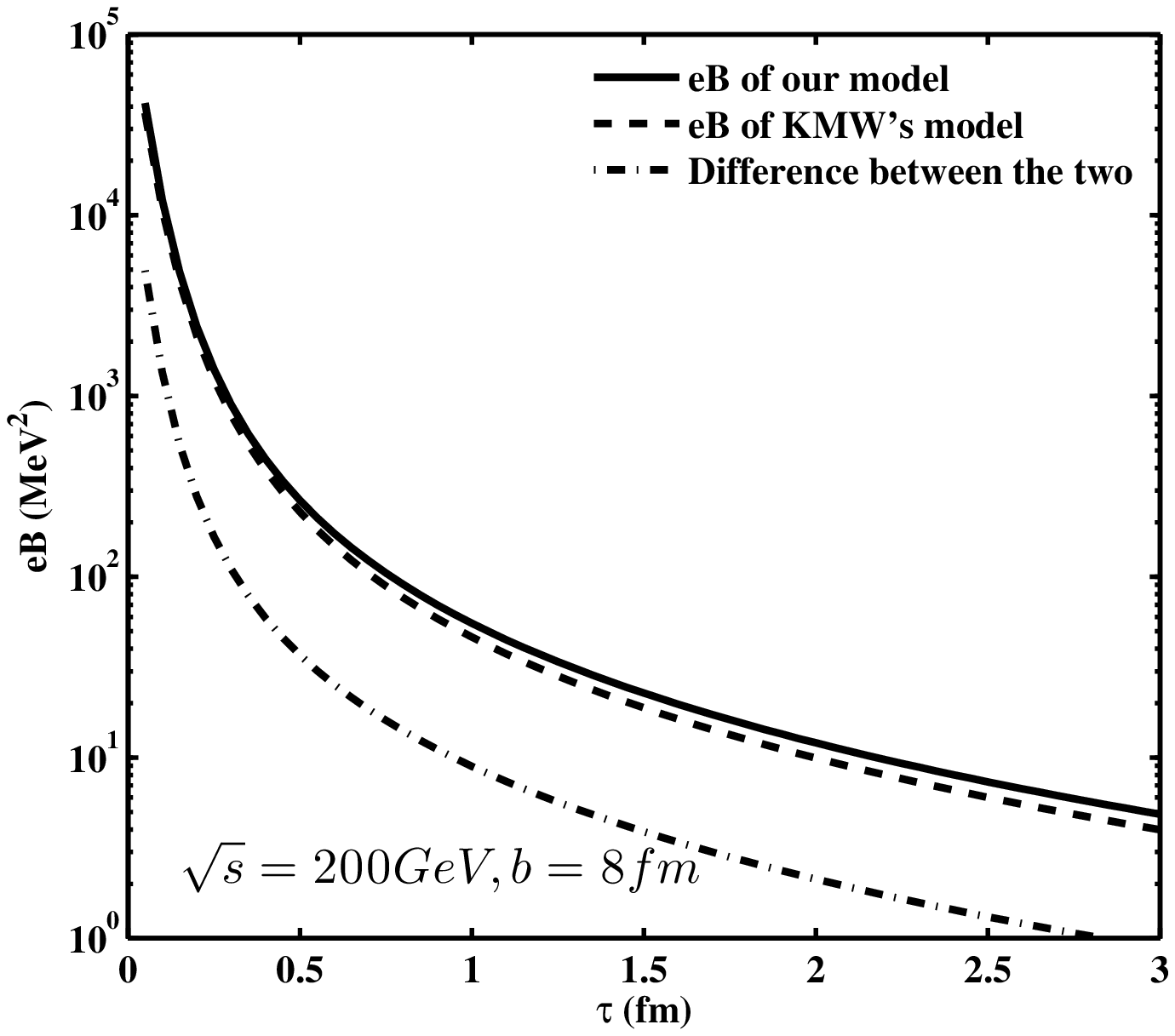}}
\caption{The same as Fig.3 but for $\sqrt{s}$ = 200 GeV.}
\label{fig4}
\end{figure}

\begin{figure}[h!]
\centering \resizebox{0.5\textwidth}{!}{
\includegraphics{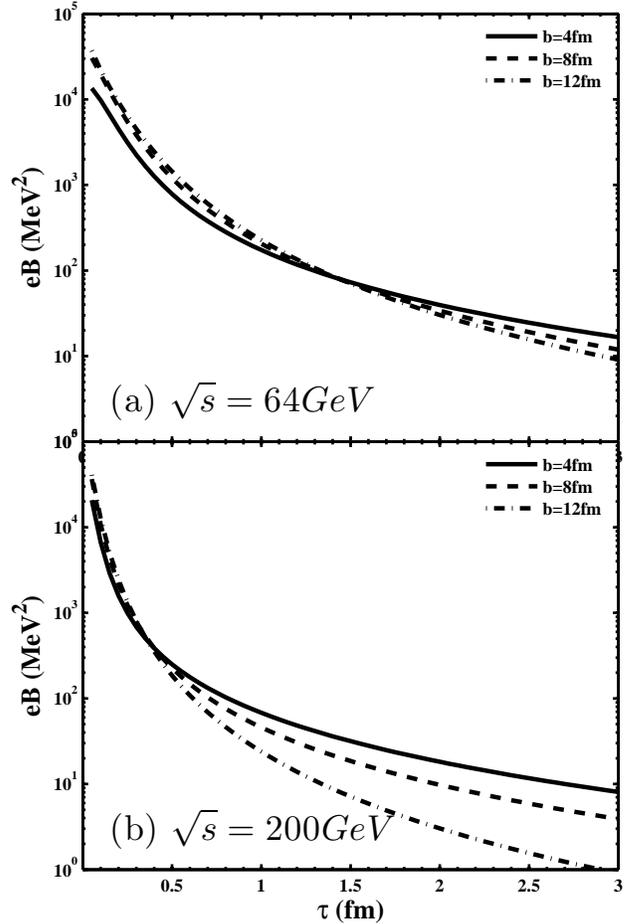}}
\caption{The dependencies of magnetic field on proper time for Au-Au collisions at different collision energies for
(a) $\sqrt{s}$ = 64 GeV  and (b) $\sqrt{s}$ = 200 GeV.}
\label{fig5}
\end{figure}

\begin{figure}[h!]
\centering \resizebox{0.50\textwidth}{!}{
\includegraphics{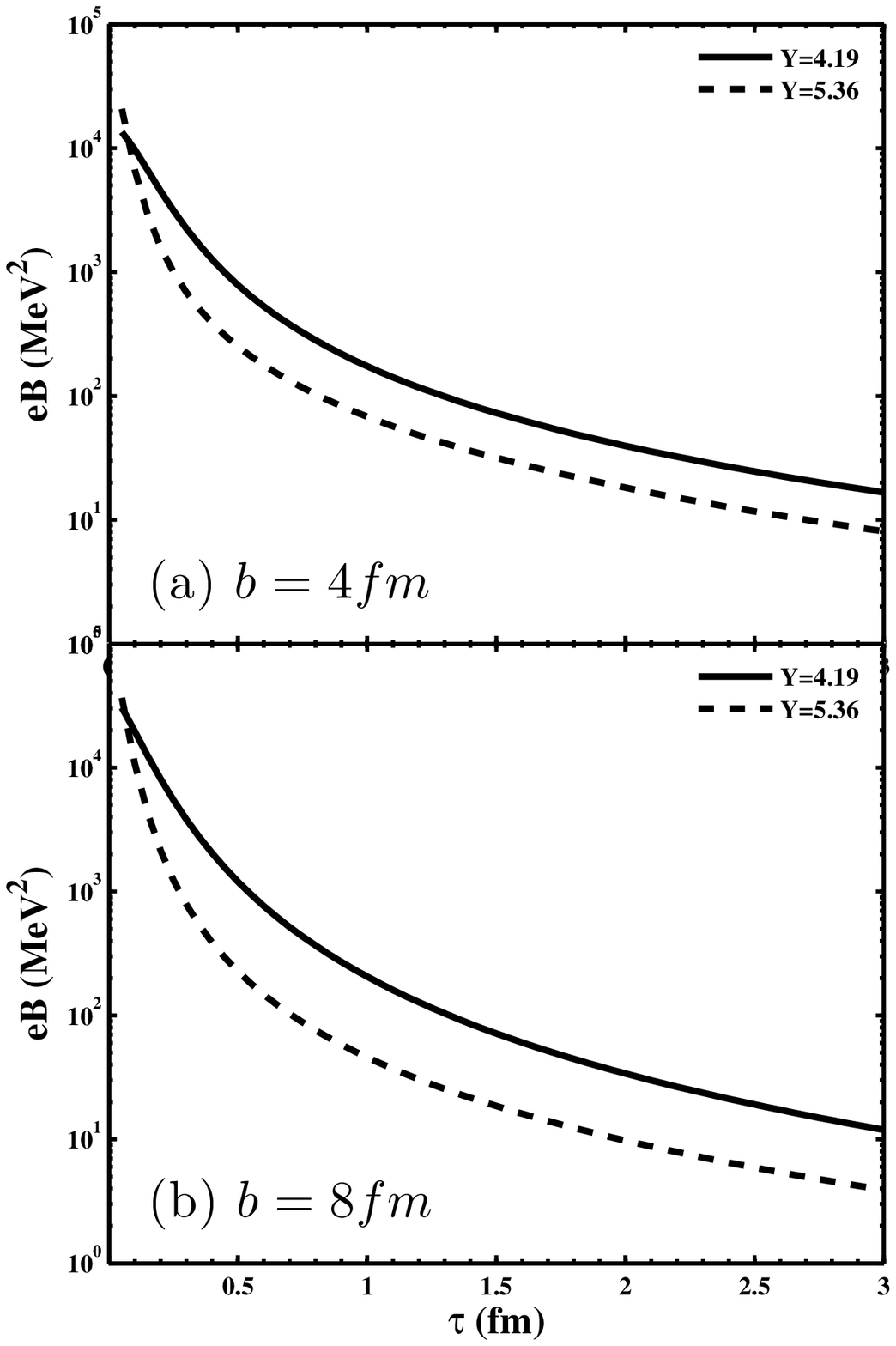}}
\caption{The dependencies of magnetic field on proper time for Au-Au collisions at different impact parameters of
(a) $b$ = 4 fm and (b) $b$ = 8 fm.}
\label{fig6}
\end{figure}

Figure 3 shows the dependence of the magnetic field on the proper time for Au-Au collision with $b$ = 8 fm at $\sqrt{s}$ = 64 GeV.
The solid line is our calculation results  using the Wood-Saxon nuclear distribution, and the dashed line denotes the KMW results  using
the uniform nuclear distribution. The dash dotted line is the difference  between our model and KMW model. From Fig. 3 we knows that the magnetic fields can indeed
be created in noncentral heavy-ion collisions and it is this field that makes it possible to separate
the right- and left-hand quarks. Figure 3 and 4 also show that the size of the field is quit large especially just after the collision and decreases rapidly over time,
and the magnitude of the magnetic field using the Wood-Saxon nuclear distribution is slightly bigger that that of by using the uniform distribution.
Figure 4 shows the dependence of the magnetic field on proper time for a Au-Au collision with $b$ = 8 fm  $\sqrt{s}$ = 200 GeV. The same situation is also shown in Fig. 3.

Figure 5 shows the dependencies of the magnetic field on the proper time for Au-Au collisions at different collision energies $\sqrt{s}$ = 64 GeV [Fig.5a]
and $\sqrt{s}$ = 200 GeV [Fig.5b], respectively.
Figure 5(a) shows us that the magnetic field is slightly large for large impact parameter $b$ when $\tau \leq$ 1.2 fm
at $\sqrt{s}$ = 64 GeV, and $\tau \sim$ 1.2 fm
is a cross point. But when $\tau$ > 1.2 fm, the magnetic field is relatively larger for small impact parameter $b$ than it is for large impact parameter.
Figure 5(b) shows that when the energy of the central-of-mass central system increases from $\sqrt{s}$ = 64 GeV to $\sqrt{s}$ = 200 GeV,
the magnetic fields are nearly unchanged when  $\tau \leq$ 0.3 fm, and  $\tau \sim$ 0.3 fm
is a cross point. The magnitude of the cross point at $\sqrt{s}$ = 200 GeV is far less than that of $\sqrt{s}$ = 64 GeV. When $\tau$ > 0.3 fm,
the magnetic field is relatively large for small impact parameter $b$.

Figure 6 shows the dependencies of the magnetic field on proper time for Au-Au collisions at different impact parameters of  $b$ = 4 fm [Fig.6a] and $b$ = 8 fm [Fig.6b], respectively.
Figure 6(a) and 6(b) show us that the magnetic field at $\sqrt{s}$ = 64 GeV is larger  than that of $\sqrt{s}$ = 200 GeV.

\begin{figure}[h!]
\centering \resizebox{0.50\textwidth}{!}{
\includegraphics{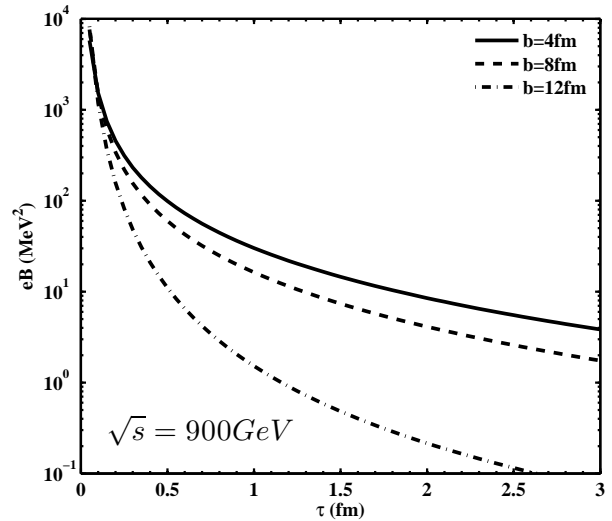}}
\caption{The dependencies of the magnetic field on proper time for Pb-Pb collisions and $\sqrt{s}$ = 900 GeV at different
impact parameters of $b$ = 4 fm,$b$ = 8 fm and $b$ = 12 fm, respectively.}
\label{fig7}
\end{figure}

~~~~~~~~~~~~~~~~~~~~~~~~~~~~~~~~~~~~~~~~~~~~~~~~~~~~~~~~~~~~~~~~~~~~

We have done further researches based on the discussion above. The magnetic field at the LHC energy regions is predicted. by using Eqs. (9) and (12).
Figure 7 shows the dependencies of the magnetic field on proper time for Pb-Pb collisions and $\sqrt{s}$ = 900 GeV at different
impact parameters of $b$ = 4,$b$ = 8 and $b$ = 12fm, respectively. Figure 7 shows that, at $\tau \sim$ 0, the magnitudes of magnetic field at $b$ = 4,$b$ = 8 and $b$ = 12 fm
are nearly same, but the magnitudes of the magnetic field decrease as  $\tau$ increases. It is shown that the magnitudes of the magnetic field of more off-central collision ($b$ = 12 fm)
drop dramatically along with the time. We also predict the dependencies of a magnetic field on proper time for Pb-Pb collisions and $\sqrt{s}$ = 2760 GeV (Fig. 8) and at $\sqrt{s}$ = 5500 GeV (Fig. 9), respectively. Figures 8 and 9 present the same rule as in Fig. 7. It is obvious that the magnitude of the magnetic field in the LHC energy region is not as big as
the ones in the Relativistic Hravy-Ion Collider (RHIC) energy regions.

\begin{figure}[h!]
\centering \resizebox{0.50\textwidth}{!}{
\includegraphics{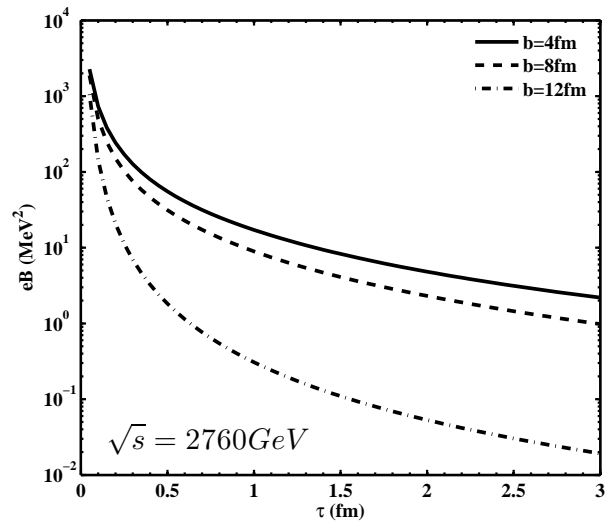}}
\caption{The dependencies of the magnetic field on proper time for Pb-Pb collisions and $\sqrt{s}$ = 2760 GeV  at different
impact parameters of $b$ = 4,$b$ = 8 and $b$ = 12 fm, respectively.}
\label{fig8}
\end{figure}

\begin{figure}[h!]
\centering \resizebox{0.50\textwidth}{!}{
\includegraphics{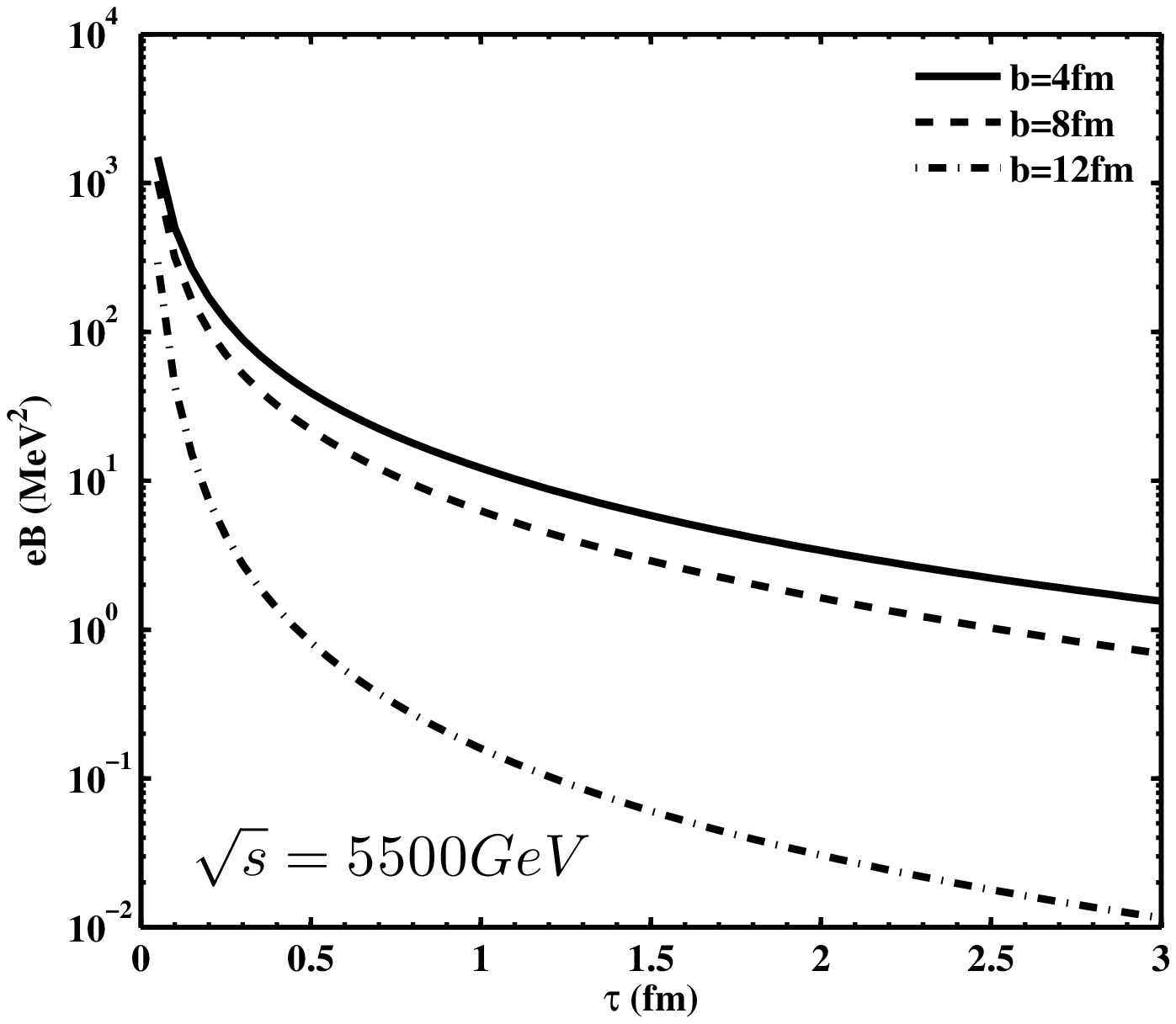}}
\caption{The same as Fig.8 but for $\sqrt{s}$ = 5500 GeV.}
\label{fig9}
\end{figure}

The analysis shows an enormous magnetic field can indeed
be created in off-central heavy-ion collisions and it is this field that makes it possible to separate
the right- and left-hand quarks. The size of the field is quite large, especially just after the collision, and decreases rapidly over time.

\section{The magnetic field of total charges by the produced quarks}
To discuss the produced quark distribution, we start with the momentum spectrum of quarks radiated by a stationary thermal source with
temperature $T_{th}$:

\begin{equation}
E\frac{d^{3}N_{th}}{d^{3}p}=\frac{d^{3}N_{th}}{{\rm d}Y{\rm d}p_{T}{\rm d}\phi}\propto{Ee^{-E/T_{th}}}
\label{eq:eq13} 
\end{equation}

In the following we give the spectra in terms of rapidity $Y=\tanh^{-1}(p_{\textrm{L}/E})$.

The rapidity distribution of thermal quarks can be given by integrating Eq. 13 over the transverse
component~\cite{lab22,lab23}, such as
\begin{equation}
\frac{dN_{th}}{dY}\propto\frac{mT_{th}}{(2\pi)^{2}}(1+2\xi_{0}+2\xi^{2}_{0}){\rm e}^{(-1/\xi_{0})},
\label{eq:eq14} 
\end{equation}
\noindent where $\xi_{0}=T_{th}/(m\cosh{Y})$.

Equations (13) and (14) give the isotropic thermal distribution. As mentioned in Refs.~\cite{lab24,lab25,lab26,lab27}, the
measured momentum distribution of the produced particles is certainly anisotropic~\cite{lab28,lab29,lab30,lab31,lab32,lab33,lab34,lab35,lab36,lab37}. It is privileged in the direction
of the incident nuclei. This is because the produced particles still carry their's kinematic information,making
the longitudinal direction more populated than the transverse ones. The simplest way~\cite{lab22,lab23,lab28} to account
for the anisotropy is to add up the contributions from a set of fireballs with centers located uniformly in the
rapidity region $[-Y_{f0},Y_{f0}]$. The corresponding rapidity distribution is obtained through changing $\xi_{0}$ into
$\xi=T_{th}/[m\cosh(Y-Y_{f})]$ and integrating over $Y_{f}$ from $-Y_{f0}$ to $Y_{f0}$:
\begin{equation}
\frac{dN_{th}}{dY}\propto\int_{-Y_{f0}}^{Y_{f0}}dY_{f}\frac{{\rm m}{T_{th}}}{(2\pi)^{2}}(1+2\xi+2\xi^{2}){\rm e}^{(-1/\xi)},
\label{eq:eq15} 
\end{equation}

\noindent where $\xi=T_{th}/[m\cosh(Y-Y_{f})]$. The distribution of produced quarks can be given by
\begin{equation}
f(Y)=K\int_{-Y_{f0}}^{Y_{f0}}\left(1+2\Gamma+2\Gamma^2\right){\rm e}^{-1/\Gamma}{\rm d}Y_{f}
\label{eq:eq16} 
\end{equation}

It also should be normalized and $K$ is a normalization constant.Here
\begin{equation}
\Gamma=\frac{T_{th}{\rm e}^{-(\tau-\tau_0)/\tau_0}}{m\cosh(Y-Y_f)}
\label{eq:eq17} 
\end{equation}

Let us try to figure out how the density of the electric charge should be dependent on the chemical potential
and the nucleon number of colliding nuclei,
\begin{equation}
\rho_{q\pm}(\vec{x}^\prime_{\bot},Y)=\kappa\frac{\rho_{\pm}(\vec{x}^\prime_\bot)}{1+{\rm exp}(\frac{\varepsilon(Y)-\mu_q}{T})}
\label{eq:eq18} 
\end{equation}

\noindent where $\varepsilon(Y)$ is the quark energy as a function of the rapidity $Y$ of produced quarks, $\mu_q$ is the quark chemical potential, $T$ is temperature, and $\kappa$ is a normalization constant. The energy of the quark is given as:
\begin{equation}
\varepsilon(Y)=\sqrt{m^{2}+p_{\rm T}^{2}}\cosh Y
\label{eq:eq19} 
\end{equation}

\noindent where we assume the average transverse momentum $p_{\rm T}$ = 0.2 GeV, the constituent quark mass $m$ = 0.308 GeV and the temperature $T$ = 0.2 GeV.
With the parameterizations of $T$ and baryon chemical potential $\mu_{B}$ from
Fig.1 of Ref.~\cite{lab38}, we assume
that the $\mu_{B}$ = 0.03 GeV with $\mu_{q}$ = 0.01 GeV at $\sqrt{s}$ = 200 GeV, and the $\mu_{B}$ = 0.06 GeV with $\mu_{q}$ = 0.02 GeV  at $\sqrt{s}$ = 64 GeV.

Then we get the expression for the magnetic field of the total charge given by the produced quarks:
\begin{eqnarray}
\begin{split}
e\vec{B}(\tau,\eta,\vec{x}_\bot)&={\pm}ZK\alpha_{EM}\int{\rm d}^2\vec{x}^\prime_\bot
\int{\rm d}Y\int_{-Y_{f0}}^{Y_{f0}}{\rm d}Y_{f}\\
&\times(1+2\Gamma+2\Gamma^2){\rm e}^{-1/\Gamma}\sinh(Y\mp\eta)\\
&\times\rho_{q\pm}(\vec{x}^\prime_{\bot},Y)\frac{(\vec{x}^\prime_\bot-\vec{x}_\bot)\times\vec{e}_z}
{[(\vec{x}_\bot^\prime-\vec{x}_\bot)^2+\tau^2\sinh(Y\mp\eta)^2]^{\frac{3}{2}}}
\end{split}
\label{eq:eq20} 
\end{eqnarray}

\begin{figure}[h!]
\centering \resizebox{0.50\textwidth}{!}{
\includegraphics{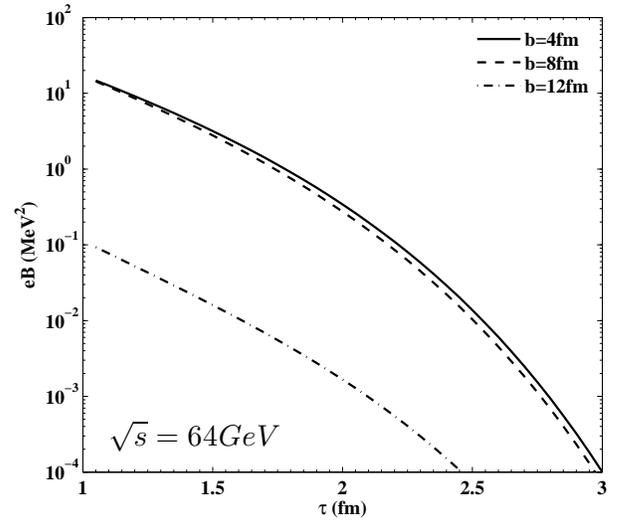}}
\caption{The dependence of magnetic field of total charge given by the produced particles on proper time for Au-Au collision with $\sqrt{s}$ = 64 GeV at $b$ = 4,$b$ = 8
 and $b$ = 12 fm, respectively.}
\label{fig10}
\end{figure}

\begin{figure}[h!]
\centering \resizebox{0.50\textwidth}{!}{
\includegraphics{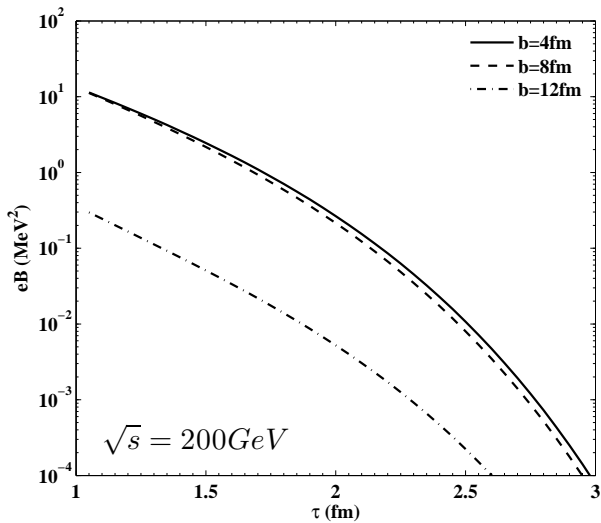}}
\caption{The same as Fig.10 but for $\sqrt{s}$ = 200 GeV.}
\label{fig11}
\end{figure}

Figure 10 shows the dependence of the magnetic field on the total charge given by produced particles at proper time for a Au-Au collision with $\sqrt{s}$ = 64 GeV at $b$ = 4,$b$ = 8
 and $b$ = 12 fm, respectively.
The solid line is for $b$ = 4 fm, the dashed line is $b$ = 8 fm, and dash-dotted line is $b$ = 12 fm.
Equation 18 is used to calculate the results of produced particles. From Eq. 18, as a whole, we can figure out that the contribution to the magnetic field
of produced particles is smaller than that of contributions of participant and spectator nucleons. a smaller of impact parameter results in a larger
magnitude of produced particles. When $b\rightarrow{12}fm$, the magnitude of the magnetic field of produced particles is approximately  $0$.

Figure 11 shows the dependence of the magnetic field on the total charge given by produced particles on proper time for Au-Au collision with $\sqrt{s}$ = 200 GeV at $b$ = 4,$b$ = 8
 and $b$ = 12 fm, respectively.

\section{Summary and Conclusion}
In this paper, we start from the magnetic field produced by traveling charged particles and estimate the magnetic field
in a reasonable way with the Wood-Saxon distribution instead of the model of uniform distribution.
We acquired useful results at the same time.

The QCD vacuum with nonzero winding number $Q_w$, which is created in heavy-ion collisions, breaks the
CP symmetry.
As a result, we obtain a difference between the number of left- and right-handed fermions for each flavor.
In the presence of a background magnetic field $B$ which is also created in heavy-ion collision,
the right and left-handed fermions  move oppositely and then a charge difference between opposite sides
of the reaction plane is induced. This is the chiral magnetic effect.
This kind of charge difference has been indeed observed in high-energy physics experiments and the phenomenon
can be used as a proof of CP violation.
We can also use the magnetic field we have and another model to estimate this kind of effect theoretically.

We show that an enormous magnetic field can indeed
be created in off-central heavy-ion collisions.  It is shown that this magnetic field  makes it possible to separate
the right- and left-hand quarks. The magnitude of the field is quite large, especially just after the collision, and decreases rapidly with time.
The drop velocity increases with the collision energy increase. It is shown that the magnitudes of the magnetic field of more off-central collision at the LHC energy region
drops dramatically along with the time. We also predict the dependencies of magnetic field on proper time for Pb-Pb collisions at $\sqrt{s}$ = 2760 GeV  and  $\sqrt{s}$ = 5500 GeV (Fig. 9), respectively.

We also show the dependence of the magnetic field of the total charge given by the produced particles on proper time for a Au-Au collision with $\sqrt{s}$ = 64 GeV and  $\sqrt{s}$ = 200 GeV  at $b$ = 4, $b$ = 8 and $b$ = 12 fm, respectively. It is shown that the contribution to the magnetic field of the total charge given by the produced quarks is smaller than that of contributions of participant and spectator nucleons.  We also find that the magnitude of the magnetic field given by produced quarks increases with the impact parameter decreases.

\section{Acknowledgments}
This work was supported by National Natural Science Foundation of China (Grant No: 10975091), the CCNU-QLPL Innovation Fund (Grant No: QLPL2011P01),
the Excellent Youth Foundation of Hubei Scientific Committee (Grant No: 2006ABB036), and the Education Commission of Hubei
Province of China (Grant No: Z20081302).The authors are indebted to
Professor Lianshou Liu for his valuable discussions and very helpful suggestions.

{}

\end{document}